\documentclass[sigconf]{cidr-2025}
\usepackage{graphicx} 
\usepackage{booktabs}
\definecolor{ForestGreen}{RGB}{34,139,34}
\definecolor{Orange}{RGB}{255, 121, 0}
\usepackage{amsmath} 
\newtheorem{listing}{Listing}[section]

\newcommand{\todo}[1]{}
\newcommand{\nice}[1]{}
\renewcommand{\todo}[1]{{\color{ForestGreen} TODO: {#1}}}
\renewcommand{\nice}[1]{{\color{Orange} NICE-TO-HAVE: {#1}}}

\usepackage{xcolor}
\usepackage{soul}
\usepackage{tikz}
\newcommand{\hlc}[2][yellow]{{%
    \colorlet{foo}{#1}%
    \sethlcolor{foo}\hl{#2}}%
}

\newcommand{\dataset}{Goby }

\title{Mind the Data Gap: Bridging LLMs to Enterprise Data Integration} 
\author{Moe Kayali}
\email{kayali@cs.washington.edu}
\orcid{0000-0002-0643-6468}
\affiliation{
\institution{University of Washington}
\country{}
}

\author{Fabian Wenz}
\orcid{0009-0008-8419-7205}
\email{fab_wenz@mit.edu}
\affiliation{
\institution{TU Munich and MIT}
\country{}
}

\author{Nesime Tatbul}
\orcid{0000-0002-0416-7022}
\email{tatbul@csail.mit.edu}
\affiliation{
\institution{Intel Labs and MIT}
\country{}
}

\author{Çağatay Demiralp}
\orcid{0009-0003-2080-0443}
\email{cagatay@csail.mit.edu}
\affiliation{
\institution{AWS AI Labs and MIT}
\country{}
}

\date{June 2024}

\begin{abstract}
Leading large language models (LLMs) are trained on public data. However, the majority of the world's data is \textit{dark data} not publicly accessible, mainly in the form of private organizational data or \textit{enterprise data}. We show that the performance of methods based on LLMs seriously degrades when tested on real-world enterprise datasets. Current benchmarks, based on public data, overestimate the performance of LLMs. We release a new benchmark dataset, the \textit{\dataset Benchmark}, to advance discovery in enterprise data integration. Based on our experience with this enterprise benchmark, we propose techniques to uplift the performance of LLMs on enterprise data, including: (1) hierarchical annotation, (2) runtime class-learning, and (3) ontology synthesis. We show that, once these techniques are deployed, the performance on enterprise data becomes on par with that of public data. The Goby benchmark can be obtained at \url{https://goby-benchmark.github.io/}.
\end{abstract}

\begin{document}

\maketitle

\section{Introduction}
Despite intensive academic and industrial interest, the uptake of large language models (LLMs) for data management and integration tasks remains limited in practice. A study by Gartner reports that at least 30\% of LLM-based enterprise projects will be abandoned after proof of concept by the end of 2025~\cite{Research:2024aa}. 
 For example, commercial LLM-based legal research tools by industry leaders LexisNexis and Thomson Reuters were measured to have only 40-65\% accuracy~\cite{DBLP:journals/corr/abs-2405-20362}. Their failures were sufficiently high profile that John Roberts, chief justice of the United States Supreme Court, rebuked lawyers using LLM-based tools in his 2023 summary of the state of the judicial system~\cite{Roberts:2023aa}.

LLM-based techniques have been intensively studied for data integration over the last two years by the database community, including entity matching~\cite{Narayan:2022ab}, column type annotation~\cite{Kayali:2024aa}, wrapper induction~\cite{DBLP:journals/pvldb/AroraYENHTR23}, and candidate key identification~\cite{Trummer:2022aa}. Industry data management practitioners have also embraced these techniques, with LLM-based techniques deployed in Azure Data Lake, Microsoft Excel, and Amazon Q for Business. Several startups in the LLM-for-data-management space, including Unstructured and Numbers Station, have attracted tens of millions in investment dollars. Interest continues to be high.

Public benchmarks are widely used to quantify LLM performance on these tasks. Examples include \textit{VizNet}~\cite{Hu:2019aa}, \textit{T2Dv2}~\cite{DBLP:conf/edbt/RitzeB17}, and \textit{FDA-510k}. These benchmarks are overwhelmingly comprised of public data collected from the web. In the VizNet dataset, 25\% of data is from open government datasets (\textit{e.g.,} World Bank and NOAA tables), and another 25\% is from public web tables. In T2Dv2, 100\% of tables are fully wiki-type tables. The most popular label in that dataset is \texttt{PopulatedPlace}, which corresponds to well-known cities and towns. In FDA-510k, the dataset is comprised entirely of public Federal Food and Drug Administration reports on the web. 

On these benchmarks, several studies find LLM-based approaches are state of the art~\cite{Kayali:2024aa}. However, the aforementioned Gartner report finds that the most common reason for abandoning LLM-based technology is the low quality of results. This posits the question: \textit{what explains the LLM performance gap between research and practice?}

These challenges are not new. An industry survey by Rexer Analytics in 2023 reports that only 32\% of machine learning implementations in the enterprise achieve deployment~\cite{Analytics:2023aa}. Among solutions that involve the introduction of fundamental architectural changes (such as LLMs), only 22\% reach successful deployment. A significant reason is population drift: the data on which the model is deployed are significantly different from the data it is trained on. An influential early work indicates that at consumer finance firms, learned approaches last two months before they are no longer effective on the shifting data distribution and must be re-trained~\cite{10.1214/088342306000000060}.

This highlights a fundamental limitation of LLMs in data integration: out-of-the-box models like GPT are likely to fail. A survey on LLM-based data annotation emphasizes that while LLMs can automate labeling processes, they often inherit biases from their training data and exhibit overconfidence in their predictions. Consequently, companies frequently rely on human oversight to ensure the accuracy and reliability of the labeled data~\cite{DBLP:conf/emnlp/TanLWBJBKL0024}.

\textit{This work is a call to action: evaluating LLM performance on public data management benchmarks risks presenting 
an overly rosy picture of its abilities.}  In the first part of the paper, Section~\ref{sec:background}, we discuss data management benchmarks, explaining that they can exist on a spectrum of public-to-private sources. Then, we present \textit{Goby}, a new benchmark built with real enterprise in Section~\ref{sec:dataset}. Finding that LLM-based approaches perform more poorly on the Goby Benchmark than comparable public-data benchmarks, we propose new concepts for adapting LLMs to enterprise data and empirically evaluate these in Section~\ref{sec:insights&Experiments}. Finally, we discuss our results and future directions in Section~\ref{sec:discussion}. \textit{To resolve this data gap, we in the data management community must ``do the work'' and develop more realistic benchmarks.}


\section{Background}
\label{sec:background}



Benchmarks and their accompanying datasets can be categorized, in general terms, into two buckets: public and private. On the fully-public end, there are standard benchmarks such as VizNet~\cite{Hu:2019aa} and GitTables~\cite{DBLP:journals/pacmmod/HulsebosDG23}. These contain verbatim public data from web sources. Another well-known example is the Spider Benchmark~\cite{DBLP:conf/emnlp/YuZYYWLMLYRZR18}, which compiles databases from sources such as Wikipedia but generates questions over them using the traditional graduate-student-based approach. On the other end, there are private datasets made from proprietary data sources and labeled with proprietary labels, which are not possible to understand outside the originating organization. Our dataset is oriented towards these private datasets.

Further, separate from the issue of public accessibility is the over-representation of some high-visibility domains in benchmarks. Examples of such domains are celebrities, government, and geography. While these domains are very well-represented on the public web, they do not make up the bulk of enterprise data.

Task suitability is another concern. Many existing benchmarks are constructed in a manner not authentic to enterprise use. These benchmarks are created by crawling the web for specific patterns that are easy to identify. For example, GitTables is constructed by searching GitHub for patterns such as \texttt{id} and \texttt{state}, and similarly, for T2Dv2, tables that were easy to annotate were selected from Wikipedia. In other words, the ``how'' of data collection dictates the ``what'' of the database content. On the other hand, organizations generally decide on ``what'' datasets they have a business need for and then the ``how'' of data collection.

Virtually all academic systems are evaluated on public datasets (\textit{e.g.},~\cite{Kayali:2024aa, Narayan:2022ab, DBLP:journals/pvldb/AroraYENHTR23}). Gartner reports that lack of ability to perform evaluation is common in enterprise setting: industry generally does not have its own benchmarks~\cite{Research:2024aa}. In personal communication with practitioners building data management systems, we have found this to be true.

Many research systems focused specifically on public data, such as the seminal WebTables system~\cite{DBLP:journals/pvldb/CafarellaHWWZ08}. Prior work noted the poor generalization of many deep learning techniques~\cite{Kayali:2024aa}. The reason for that is the high risk of overfitting to the training data, which restricts the model's ability to generalize and adapt to new, unseen data and contexts. For example, consider the performance of the deep learning model DoDuo~\cite{Suhara:2022aa}. In the setting where it is trained and tested on the same data distribution (VizNet), the performance is strong. However, when we attempt to generalize by training on the similarly structured Wiki dataset and testing it on VizNet, its performance drops by at least 30\% in terms of F1-score, precision, and recall~\cite{Kayali:2024aa}. This comparison is merely between two public datasets with very similar data types and structures. An unseen private dataset, which often has unique structures and proprietary labels, may pose an even bigger challenge. This underlines the necessity for more robust approaches to bridge the gap between private and public data sources.

We focus this paper on a specific data integration task: semantic column type annotation~\cite{DBLP:conf/kdd/HulsebosHBZSKDH19}. Given a relational table, column annotation involves assigning to each column a type which corresponds to the type of values stored in that column. The classes are drawn from a set of labels: if those labels are basic types (e.g., date, time), the problem is called \textit{atomic} type annotation, while if the labels have semantic meaning (e.g., birthday, deadline date, best-by date), it is \textit{semantic}. We focus on the latter formulation in this paper. 



\section{\dataset Benchmark}
\label{sec:dataset}
\textit{\dataset} is a benchmark dataset we derive from a production industry workload in the event promotion and marketing setting. The dataset was compiled around 2017 and encompasses over 4 million events.

\paragraph{Constructing \dataset} The dataset was constructed over several years. First, over one thousand wrappers were written by professional developers in order to convert web pages and APIs into relational tables. Once those wrappers were developed, they were executed to obtain 4.1 million rows of data, each corresponding to an event. Next, the team developed a set of semantic types common to all the events. The labels were applied to ~40\% of the columns. The rest of the columns were miscellaneous data not of interest to the task. Finally, the semantic type mappings were used to create a universal schema which unifies all the source tables into one.

\paragraph{Benchmark components} Its components are: 

\begin{enumerate}
    \item a semantic type hierarchy developed by domain experts
    \item nearly 1~200 source tables, each corresponding to the output of one wrapper
    \item mapping each annotated with the semantic types of its columns
    \item a universal table in which all data from the source tables is unified
\end{enumerate}

\paragraph{Benchmark Characteristics} All in all, \dataset is comprised of a total of 4.04 million rows and 23~203 columns.  Key characteristics quantifying the dataset are provided in Table~\ref{tab:goby-char}. 

\dataset exhibits high task suitability for data integration tasks. This is because the construction method is realistic and tailored for a data integration pipeline. We also highlight that \dataset is semantically-rich as compared to with other benchmarks. The most common labels are those related to events, such as location and organizer details. Meanwhile, typical benchmarks contain labels with low semantic information, for example the three most popular semantic types in one benchmark~\cite{DBLP:journals/pacmmod/HulsebosDG23} are the generic labels \texttt{id}, \texttt{name} and \texttt{type}.

Further, because of the method of construction, \dataset has more realistic table structure. This can be seen in the average number of rows: 3~400, and columns: 20. This reflects the need to have a minimum amount of business information about each event. Prior benchmarks such as VizNet have an average of only 3 columns per dataset, while T2Dv2 has only 118 rows per table; both are unrealistic. The larger number of rows and columns reflects the reality of enterprise data. 

\begin{table}
  \centering
  \caption{Key characteristics of the \dataset dataset.}
    \begin{tabular}{lr}
        \toprule
        \textbf{\dataset} & \\
        \midrule
          \# Tables                    & 1~187                \\
          Avg. Rows / Table                    & 3~405                \\
          Avg. Col / Table                    & 20                \\
          Total Rows                    & 4~042~519              \\
          Total Columns                    & 23~203               \\   
        \bottomrule
    \end{tabular}
  \label{tab:goby-char}
\end{table}

While some of the underlying data from \dataset has been used in a prior data curation paper~\cite{Stonebraker:2013aa}, we now curate and release the dataset as a benchmark for the first time.

\dataset is a private benchmark, according to the public-private taxonomy presented in section~\ref{sec:background}. In particular, while the domain doesn't require proprietary knowledge (events around town), the data is not from web tables but scrapped with human-programmed scripts from APIs, then labelled with private labels. To the best of our knowledge, \dataset is the first data integration benchmark derived from a private dataset.


\dataset does not contain any private personal identifiable information (PII). It also does not contain any data pertaining to individual users.

\textbf{Download \dataset at \url{https://goby-benchmark.github.io/}.}

\begin{figure}
    \centering
    \includegraphics[width=1.05\linewidth]{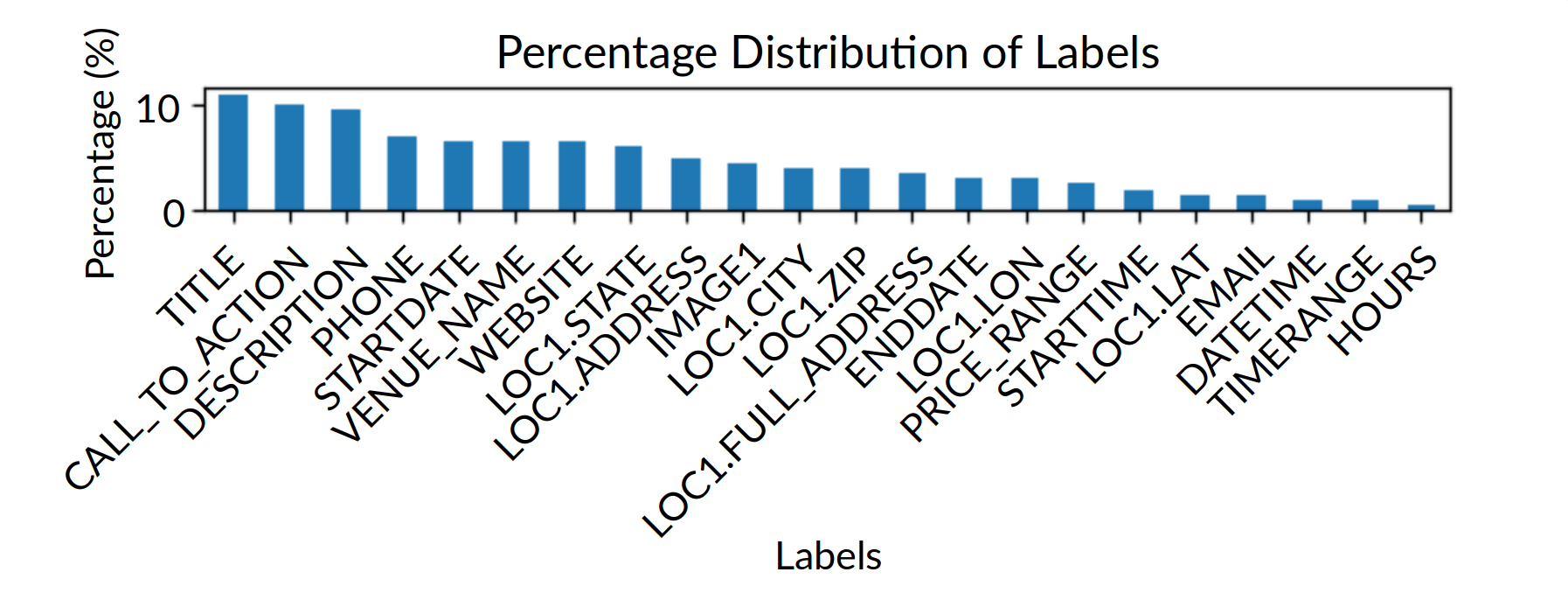}
    \caption{Semantic column label distribution in the \dataset dataset.}
    \label{fig:goby-labels}
\end{figure}

\section{Insights and Experiments}
\label{sec:insights&Experiments}
In this section, we explore the performance of large language models (LLMs) in the context of enterprise data, highlighting the limitations of previous models and benchmarks, as well as novel approaches we discovered for optimizing LLM performance. Through our experiments, we present key insights that not only build upon prior work but also introduce innovative techniques for integrating and analyzing enterprise data using LLMs. To backup these observations, we provide empirical evidence and detail in the design decisions made during the model development process, offering a clear reasoning for each step.

\paragraph{Experimental setup} We conduct an experimental evaluation using the following large language models: OpenAI's \texttt{gpt-3.5-turbo} and Meta's \texttt{Llama2} with 13 billion parameters. Experiments utilizing grammar-constrained decoding were conducted using \texttt{Llama2} since OpenAI \textsc{api}s do not expose the required raw token probabilities.

\subsection{Enterprise performance gap}
We begin by evaluating LLM approaches on the \dataset enterprise dataset in comparison to the public benchmark VizNet for the task of semantic column-type annotation. Overall, we find the outcome that LLMs perform considerably worse on \dataset compared to the public data benchmark.

The semantic column-type annotation task is a multi-class classification problem. We evaluate using the standard metrics precision $P$, recall $R$ and their harmonic mean, the $F_1$ score. Precision is defined as ratio of true positives to the sum of true positives and false positives. Recall is defined as the ratio of true positives to the sum of true positives and false negatives. The $F_1$ score reflects a metric in which precision and recall are equally weighted. Because of the multi-class setting, the reported scores are macro-averaged. Additionally, we note that the LLM can decline to answer or provide an unparsable answer, lowering the recall but not precision.

The underlying approach for extracting semantic column-type predictions is the same as the \textsc{Chorus} system~\cite{Kayali:2024aa}, we refer the reader to that paper for more details. In this experiment, the model is given access to the set of valid labels. We will remove this assumption in a later experiment.

Empirically, we find a gap between prior benchmarks and Goby of $0.18$ $F_1$ points, $14.1$ percentage points of precision and $18.8$ precentage points of recall. This is shown in Table~\ref{tab:grav}.

\begin{table}
  \centering
  \caption{Performance discrepancies in LLM performance for column-type annotation on our enterprise benchmark Goby versus the public benchmark VizNet.}
    \begin{tabular}{lccc}
        \toprule
          Metric                &  \textbf{Goby}             & VizNet~\cite{Kayali:2024aa}             & Difference           \\
        \midrule
          Precision $P$  & 77.1\%  & 91.2\%     & 14.1\%              \\
          Recall $R$       & 70.0\%   & 88.8\%   & 18.8\%             \\
          $F_1$ Score    & 0.71   & 0.89          & 0.18             \\
        \bottomrule
    \end{tabular}
  \label{tab:grav}
\end{table}

\subsection{Iterative dictionary construction}

The first step in our approach involves synthesizing a data dictionary and then generalizing it into an ontology. A Data dictionary is a list of possible data types and acceptable values, optionally with some metadata on their semantic meaning. Ontologies can be considered to be data dictionaries with a hierarchy.

The development of data dictionaries involve making many arbitrary distinctions. Label granularity is an example: the curator may choose to have the classes \texttt{street address}, \texttt{full address} and \texttt{PO box} or combine them under just one heading, \texttt{address}. Further, because of coordination failure between multiple labelers, redundant synonymous labels such as \texttt{wages} and \texttt{salary} can be present. As such, finding a perfect match for the dictionary created by a human curator is \textit{not} the goal. 

Now we start testing the ability of large language models  to iteratively construct the class labels. In this experiment, we randomly sample from the Goby tables to learn the classes. The process is seeded with a small number of starting classes. We then consider a column at a time, presenting a sample of 5 unique values from that attribute. If the LLM predicts an existing class for the column, we continue onto the next column. Otherwise, we extract a new label for the column and add that to the pool of labels. The process is terminated when the quiescence condition is met: the observation of 5 tables sequentially, all of which do not require the creation of any new labels. 

\begin{table}
    \centering
    \caption{Ability to reconstruct classes. This table of the percentage of classes were we able to infer, as compared to the number of classes in the ground truth labels.}
    \vspace{0.125in}
    \begin{tabular}{ll}
        \toprule
        \textbf{True Label} & \textbf{Predicted Label} \\
        \midrule
        \textbf{Exact Match (71\%)} & \\
        LOC1.ADDRESS & Address    \\
        HOURS & OperatingHours \\
        LOC1.LON & Geolocation \\
        ... & \\
        \midrule
        \textbf{Semantic Match (20\%)} & \\
        STATE & Location \\
        ZIP & Location \\
        ... & \\
        \midrule
        \textbf{No Match (9\%)} & \\
        DESCRIPTION & Activity \\
        ... & \\
        \bottomrule
    \end{tabular}
    \label{tab:class-reconstruction}
\end{table}

The evaluation metric here is \textit{coverage}, the portion of ground-truth label for which there exists a synthesized label. We find that the LLM can obtain \textbf{70.5\% coverage}. This is shown in Table~\ref{tab:class-reconstruction}. In the 29.5\% of classes that are not recovered, 20\% are due to incorrect granularity and 10\% are semantically wrong. This motivates the next experiment to recover from granularity failures.

\subsection{Building a Hierarchy}
\label{sec:hier}

\begin{figure}
    \centering
\includegraphics[width=0.75\linewidth]{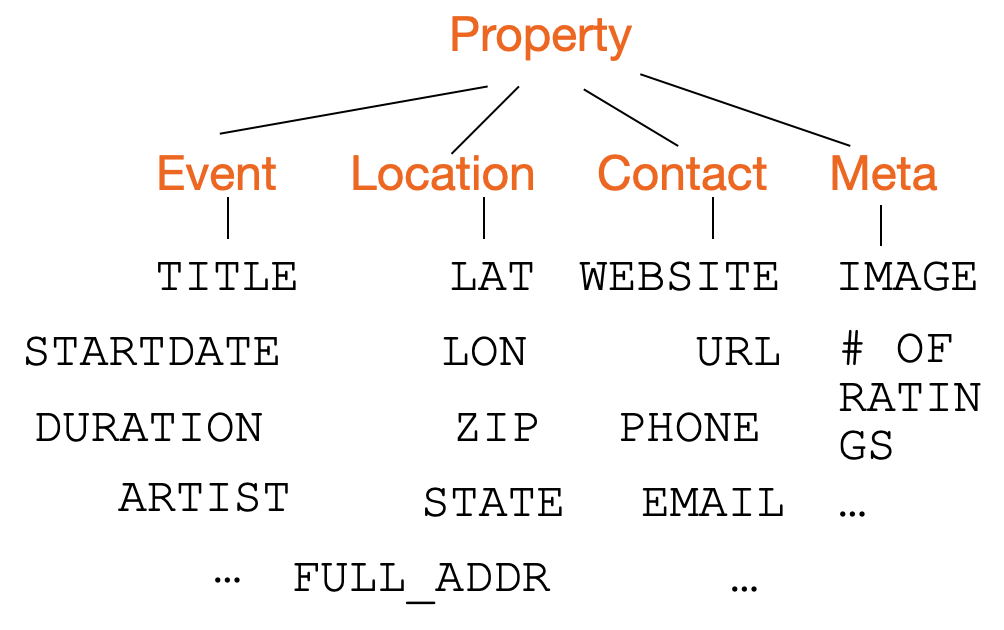}
    \caption{Example of the synthesized ontology. The base classes, in black, are first learned iteratively. The super-classes, in orange, and their relations are built in the ontology-learning step.}
    \label{fig:exmaple-ontology}
\end{figure}

We address the challenge of label granularity by learning an ontology: a structure of classes. Starting with the data dictionary of the relevant classes, we aim at the goal of creating an ontology. We consider this ontology successful if it faithfully captures all ground-truth, human-labeled classes. We show an example in Figure~\ref{fig:exmaple-ontology}. In section~\ref{sec:hier}, we will show that such a synthesized ontology can capture $100\%$ of the ground-truth labels and resolves the granularity issue.

First, we test the LLM's ability to transform its predicted classes into a hierarchy. We build the tree in an iterative approach, similar to the class learning prior. The main difference is that we utilize batching to give the model enough context to select appropriate superclasses. We find this a resounding success: the model can produce superclasses that have 100\% coverage of both the learned and ground truth classes. We show the learned ontology in Figure~\ref{fig:exmaple-ontology}. We also show the mapping from ground truth labels to the ontology in Table~\ref{tab:mapping}.

Now that we have the ontology, integrating a hierarchical approach into LLM requires a number of architectural decisions that we experiment with in the next sub-section, section~\ref{sec:serialization}. As we will see, how the hierarchy is integrated into the LLM makes a large impact on the performance.

\subsection{Encoding: Full context required}
\label{sec:serialization}
Once the ontology is learned, it must be encoded into the language model for prediction. We investigate several methods of integrating hierarchical information into LLMs. These included both architectural and semantic approaches. We consistently found that approaches that presented the full context to the LLMs outperformed those which attempted to feed only the relevant context.

\begin{figure}
    \centering
    \begin{tikzpicture}[scale=0.7, transform shape]
    \tikzstyle{input} = [rectangle, draw, fill=blue!20, text centered, rounded corners, minimum height=2em, minimum width=4em, font=\Large]]
    \tikzstyle{lm} = [rectangle, draw, fill=yellow!50, text centered, rounded corners, minimum height=2em, minimum width=4em, font=\Large]]
    \tikzstyle{output} = [rectangle, draw, fill=orange!20, text centered, rounded corners, minimum height=2em, minimum width=4em, font=\Large]]
    \tikzstyle{greenrect} = [rectangle, draw, fill=green!20, text centered, minimum height=2em, minimum width=4em]
    \tikzstyle{redrect} = [rectangle, draw, fill=red!20, text centered, dashed, minimum height=2em, minimum width=4em]
    
    \node[input] (input) at (-1, 4) {prompt};
    
    \node[lm] (lm) at (3.5, 4) {LLM};
    
    \node[output] (output) at (8, 4) {label};
    
    \node at (0, 2.5) {\textbf{Grammar-constrained decoding (GCD):}};
    
    \node at (-2, 2) {$t = 0$};
    \node at (-2, 1) {$t = 1$};
    \node at (-2, 0) {$t = 2$};
    \node at (-2, -1) {$t = 3$};
    \node at (1, -1) {...};
    
    \node[greenrect] (property) at (3.5, 2) {property};
    
    \node[greenrect] (1) at (0, 1) {event};
    \node[greenrect] (2) at (1.5, 1) {location};
    \node[greenrect] (3) at (3, 1) {contact};
    \node[greenrect] (4) at (4.5, 1) {misc};
    \node[redrect] (5) at (6, 1) {bees};
    \node[redrect] (6) at (7.5, 1) {...};
    
    \node[greenrect] (a) at (1, 0) {lat};
    \node[greenrect] (b) at (3, 0) {long};
    \node[redrect] (c) at (5, 0) {apple};
    \node[redrect] (d) at (7, 0) {...};

    \draw[->] (input) -- (lm);
    \draw[->] (lm) -- (output);
    \draw[->] (lm) -- (property);
    
    \draw[->] (property) -- (2);
    \draw[->] (2) -- (a);
    
\end{tikzpicture}
    \caption{Mechanism of grammar-constrained decoding (GCD). GCD is applied to ensure the LLM complies with the ontology to find for given data instance a correct hierarchical-classes. Possible classes are constrained to the given dictionaries and relations are based on the previous layer. During decoding only valid token continuations compliant with the grammar (shown in green) are considered. Tokens outside the grammar (shown in red) are rejected.}
    \label{fig:gcd}
\end{figure}
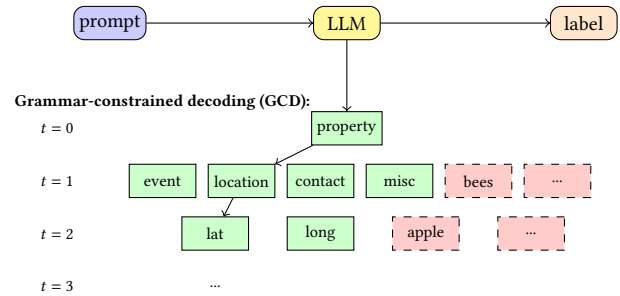

\paragraph{Grammar-constrained decoding} Large language models perform sampling to extract tokens from the conditional token probability distributions they output~\cite{Holtzman:2020aa}. Common approaches include beam search, top-$k$, temperature and nucleus sampling. This architecture leads to the possibility of constraining the search space of the sampling to match some grammar~\cite{DBLP:conf/naacl/TrombleE06}. Figure~\ref{fig:gcd} shows the mechanism by which this approach works.

\begin{listing}[Sample GBNF Grammar used for GCD]

\begin{verbatim}

root ::= answer
answer ::= "Semantic-type::" property
property ::= event | location | contact | misc
event ::= "EVENT::" sub-event
sub-event ::= "TITLE" | "STARTDATE" | "DURATION" ...
location ::= "LOCATION::" sublocation
sublocation ::= "LAT" | "LONG" | "ZIP" | "CITY" | "STATE" ...
contact ::= "CONTACT::" sub-contact
sub-contact ::= "WEBSITE" | "URL" | "PHONE" | "EMAIL" ...
meta ::= "META::" sub-misc
sub-meta ::=  "RATING" | "TOTAL_NUMBER_OF_RATINGS" ...
\end{verbatim}
\end{listing}

\paragraph{Step-by-step prompting} Another method to encode the ontology is presenting a series of prompts within the same context. This is illustrated in figure~\ref{fig:step-by-step}. Starting at the root of the class tree, the model is presented with the first level of the ontology. Once a super-class is chosen, the children of that node are presented to the model in the next prompt. This process is repeated until the model selects a leaf node. This approach can be considered a chain-of-thought method.

\begin{figure}
\centering
\fbox{\parbox{0.75\linewidth}{\textbf{Legend}: \hlc[yellow!30]{Instruction}, \hlc[cyan!30]{Data sample}, \hlc[orange!30]{Metadata}, \hlc[brown!30]{Task-specific knowledge}, \hlc[pink!30]{Prefix}.}}

\vspace{0.25in}
\raggedright
    \texttt{\hlc[yellow!30]{You are a helpful data analysis assistant. Suggest a semantic type from these classes. Reply with a single class.}}

\hlc[brown!30]{CLASSES:}

\hlc[brown!30]{PROPERTY::EVENT\_INFO}

\hlc[brown!30]{PROPERTY::EVENT\_INFO::TITLE}

\hlc[brown!30]{PROPERTY::EVENT\_INFO::DURATION}

\hlc[brown!30]{...}

\hlc[brown!30]{PROPERTY::CONTACT\_DETAILS}

\hlc[brown!30]{PROPERTY::CONTACT\_DETAILS::PHONE\_NUMBER}

\hlc[brown!30]{...}

    \hlc[cyan!30]{TABLE: serialized\_table}
    
    \hlc[orange!30]{COL-NAME:`col\_name`}
    
    \hlc[cyan!30]{VALUES: \texttt{sampled values}}
    
    \hlc[pink!30]{SEMANTIC-TYPE:}
    \caption{Tree-serialization approach. This figure shows the model prompt when serializing the ontology tree and presenting it to the model as one structure.}
    \label{fig:tree-serialization}
\end{figure}

\paragraph{Tree serialization} The final approach we consider is serializing the class tree and providing it to the model. This involves traversing the tree in pre-order and enumerating each path from the root. This allows the model's attention mechanism to selectively attend to more valid paths in the tree. Surprisingly, as will will show shortly, this approach has the most robust performance.

\begin{figure}
    \centering
    \includegraphics[width=\linewidth]{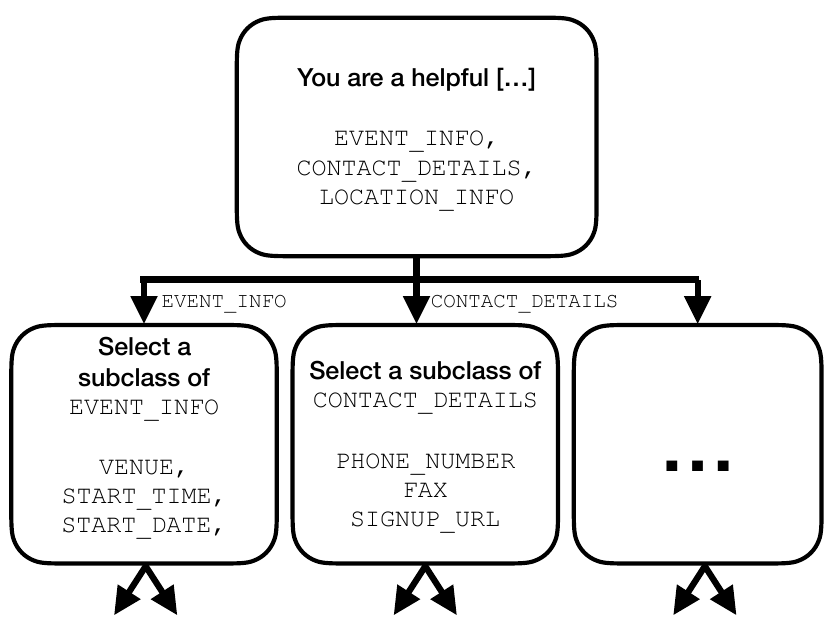}
    
    \caption{Step-by-step prompting. The model navigates a tree of potential prompts by making completions at each node. This figure shows the sequence of prompts used in this mode of operation, allowing the LLM to reason step-by-step. The prompt structure and table serialization are in line with those shown in Figure~\ref{fig:tree-serialization}.}
    \label{fig:step-by-step}
\end{figure}

\paragraph{Evaluation} We consider a prediction correct if it matches the ground truth label in our mapping, or if the most-recent common ancestor (\textsc{mrca}) is the direct parent of the prediction. We find the tree-serialization approach most effective, bringing total $F_1$ score to 0.85. Grammar-based decoding paradoxically results in a lower score $F_1$ 0.66, while step-by-step approach achieves a modest lift of 0.05 $F_1$ points from the baseline of $0.71$. This is in line with prior work that finds the GCD produces sequences that have overall low probability~\cite{DBLP:journals/corr/abs-2405-21047}. We note that the commonality between the two lower-performing approaches is that the LLM cannot examine the whole ontology: in the step-by-step prompts only one path is examined (no backtracking) while in the GCD-approach only nodes the beam search examines are considered. The approach that loads the whole ontology into the context performs best.



\begin{table}
  \centering
    \caption{Mapping the ground-truth classes into our ontology. This table shows that our ontology can accommodate all the ground-truth labels, resolving the granularity issue illustrated in Table~\ref{tab:class-reconstruction}. Overall, we achieve 100\% coverage of the classes.}
    \begin{tabular}{ll}
        \toprule
        Learned Mapping & Ground Truth Label \\
        \midrule
        PROPERTY::LOCATION::FULL\_ADDR & LOC1.ADDRESS     \\
         & LOC1.STREET\_ADDRESS          \\
         PROPERTY::CONTACT::URL & CALL\_TO\_ACTION\_URL          \\
        & WEBSITE        \\
        & OTHER\_URL           \\
        & VENUE\_LINK       \\
        & MORE\_INFO        \\
        ... & \\
        \bottomrule
    \end{tabular}
      \label{tab:mapping}
\end{table}

\section{Discussion}
\label{sec:discussion}
Our findings support the hypothesis that enterprise data presents new challenges to LLMs not captured by public-data benchmarks. We find that a hierarchical approach allows for successfully learning labels and making predictions on enterprise data.

\paragraph{Dataset contamination or distribution shift?} Two principal mechanisms may explain the observed performance gap. The \textit{data contamination} theory posits that the LLMs have been trained on (portions of) test set of the benchmark dataset. This is because the massive corpuses they are trained on are composed largely of web data, which may include the downstream test datasets. Further, LLM have the demonstrated ability to ``memorize'' discrete sequences (such as individual addresses or phone numbers) and recall them on demand---albeit only when these sequences are duplicated many times in the corpus, at least 10-100 occurrences~\cite{DBLP:conf/icml/KandpalWR22}. This contamination has been confirmed for a number of benchmarks. For instance, the MMLU (Massive Multitask Language Understanding) benchmark was also being tested and GPT-4 was able to accurately guess missing options in the benchmark test with 57\% exact accuracy~\cite{DBLP:conf/naacl/Deng0TGC24}.

Another explanation is dataset shift. As explained earlier, this involves a change in the generative distribution underlying the training and runtime populations. These models also struggle on recall of tail entities, averaging 6-20\% accuracy across all domains~\cite{DBLP:conf/naacl/SunXZLD24}. For tail entities in specialized domains, such as literature or academia, accuracy on tail entities drops to 3-5\% range~\cite{DBLP:conf/naacl/SunXZLD24}. This is evidence against the LLM memorizing benchmark answers for these more specialized benchmarks. Other evidence in support of the data shift cause is recent work which tests LLM performance on data drawn from the same distribution as the benchmarks but produced after model training. On this sub-population which temporally cannot be subject to data contamination, the authors find minimal change in model performance~\cite{Kayali:2024aa}. 


\paragraph{Cost} This also remains an elusive facet of LLM applications. Generally, the volume of ``dark'' enterprise data dwarfs that which is publicly available. Findings that LLMs are too cost prohibitive to run on public data are magnified in the enterprise setting, particularly due to extensive data duplication in the business setting. Effective table summarization becomes key: it is necessitated by the limited context window size of language models, which causes an inability to ingest whole tables. Current approaches for summarization are ad-hoc---a principled approach would be welcome. Scalability remains a unsolved challenge too. The superior performance of the (relatively context-consuming) tree serialization approach raises concerns.



The insights gleaned in this work translate more widely into general data management problems: we have a work-in-progress on applying lessons learned in this approach to problem outside data integration, focusing on natural language to query translation in the enterprise.

\section{Conclusion}
In this paper, we introduce \textit{Goby}, the first data integration benchmark utilizing real enterprise data. We show that LLM-based approaches to data integration tasks such as semantic column type annotation, having been trained on public data, experience a significant drop in performance on enterprise data (i.e., evaluating them on public benchmarks paints a false picture). More specifically, we show that out-of-the-box LLM performance on the Goby Benchmark dataset is lower than that on public datasets such as VizNet. We then propose and test approaches to bridge the gap between publicly trained LLMs and their use for enterprise data integration, primarily based on hierarchical clustering. We find these approaches effective in remedying the discovered gap for the semantic column type annotation task and invite our community for further research using on our Goby Benchmark which we are making publicly available.

\balance

\begin{acks}
We thank Michael Stonebraker and Peter Baile Chen for their valuable contributions and feedback. We thank Daniel Bruckner for assistance with the Goby data. This research is supported by Intel as part of the DSAIL program at MIT. Moe Kayali was partially supported by NSF-BSF 2109922.
\end{acks}

\bibliography{bibliography}
\bibliographystyle{plainnat}

\end{document}